\documentclass{pasj00}
\onecolumn

\begin{document}
\SetRunningHead{K. Watarai}{Light curve asymmetry of accretion flows}
\Received{2005/4/15}
\Accepted{2005/7/11}

\title{Eclipsing Light-Curve Asymmetry for Black-Hole Accretion Flows}


%
 \author{%
   Ken-ya \textsc{Watarai}\altaffilmark{1,3},
   Rohta \textsc{Takahashi}\altaffilmark{2,3}, 
   and
   Jun \textsc{Fukue}\altaffilmark{1}}
 \altaffiltext{1}{Astronomical Institute, Osaka-Kyoiku University,
 Kashiwara, Osaka 582-8582, Japan}
 \altaffiltext{2}{Graduate School of Arts and Sciences, University of Tokyo, Tokyo 153-8902, Japan} 
 \altaffiltext{3}{Research Fellow of the Japan Society for the Promotion of Science}
 \email{watarai@cc.osaka-kyoiku.ac.jp}


\maketitle

\begin{abstract}
We propose an eclipsing light-curve diagnosis for black-hole accretion flows. 
When emission from an inner accretion disk around a black hole is occulted by a companion star,
 the observed light curve becomes asymmetric at ingress and egress on a time scale of 0.1-1 seconds. 
This light-curve analysis provides a means of verifying the relativistic properties
 of the accretion flow, based on the special/general relativistic effects of black holes. 
The ``skewness'' for the eclipsing light curve of a thin disk is $\sim 0.08$,
 whereas that of a slim disk is $\sim 0$, since the innermost part is self-occulted
 by the disk's outer rim.
\end{abstract}

\section{Introduction}

Several methods have been proposed for obtaining physical information 
 about conditions near black holes.  
Direct imaging of black holes is an extremely promising method of investigation. 
For example, {\it VSOP2} and {\it MAXIM} projects are currently underway, 
and one aim of these projects is the direct imaging of black hole shadows
 (see also Hirabayashi et al. 2005; Takahashi 2005;
 and the {\it MAXIM} Web page http://maxim.gsfc.nasa.gov/).  
In particular, Sgr~A* and M87 are good targets for observing a black hole shadow 
because of their proximity and large apparent sizes.  
However, even with the use of {\it VSOP2} or {\it MAXIM},
 direct imaging studies of stellar-mass black holes
 are difficult because the size of the emitting region is extremely small.
The characteristic size of the emitting region is roughly the radius of the black hole. 
If we assume a non-rotating black hole, then the Schwartzchild radius,
 $r_{\rm g}$, would be $2.95 \times 10^6 (M/10M_\odot)$ cm. 
The radius of maximum temperature for a standard thin accretion disk is
 then located at $\sim 3 r_{\rm g}$, and therefore the size of the inner emitting region
 is $\sim$ 100 km for a $10M_\odot$ black hole. 
Therefore, observations on such small scales are extremely difficult even
 for proposed future missions. 
Consequently, timing analysis or spectroscopic study are currently more useful methods
 for investigating stellar mass black holes than the imaging studies. 
Timing analysis of quasi-periodic oscillations (QPOs) is a very popular and powerful tool 
 for examining Galactic black hole candidates (van der Klis 2000, and references therein). 
The QPO frequency can be used to derive the physical parameters of black holes, 
 i.e., the black hole mass and spin (Abramowicz \& Klu\'{z}niak 2001). 

In this paper, we propose a new method to detect accreting gas falling into a black hole
 using light curves obtained during eclipse by a companion star. 
Light curves obtained at the time of an eclipse contain information about 
 the region around a compact star, 
 that is, the curvature of the space time and angular momentum of the compact star. 
Therefore, provided the observational instrument has sufficient time resolution and sensitivity, 
 these data can be used to constrain the black hole physics. 
The basic idea for this light-curve analysis was first proposed by Fukue (1987).  
In this paper we will discuss the idea and describe methods for statistical analysis. 
Our analysis is a first step towards an ``eclipse mapping'' method (Horne 1985),
 a well-known technique for light-curve analysis of cataclysmic variables. 
This light-curve analysis method is potentially a strong tool for studying
 and obtaining physical information about black holes.  

In the next section, we examine several timescales during an eclipse in a binary system. 
In section 3, we present the results of our light-curve calculations for the eclipsing period. 
In section 4, we discuss, from various viewpoints,
 the feasibility of observationally detecting asymmetry in light curves.
The final section contains our concluding remarks.

\section{Time Scales During an Eclipse}

A light curve obtained during the eclipse of an accretion disk by a companion star contains
 physical information about the accreting gas near the black hole:
 for example, the position of the marginal stable circular orbit,
 the transonic nature of the flow, and details on radiation processes. 
This physical information is likely to provide important clues as to the nature of black holes. 
Therefore, light-curve analysis can be a strong tool for studying black holes. 
A rough estimate of the eclipsing time was derived by Fukue (1987);
 the eclipsing time scale of ingress and egress is $\Delta t \sim 2r/v_{\rm orb}$,
 where $r$ is the disk radius with an asymmetrical emission distribution, 
and $v_{\rm orb}$ is the orbital velocity. 
\begin{eqnarray}
\Delta t &\sim & \frac{2 r}{v_{\rm orb}}  \nonumber \\
         & =   & 0.85 \left(\frac{r}{20 r_{\rm g}}\right) \left(\frac{a}{R_{\odot}}\right)^{1/2}
 \left(\frac{M}{10M_{\odot}}\right) \left(\frac{M+m}{10 M_{\odot}}\right)^{-1/2}  {\rm s} \nonumber \\  
         & =   & 4.24 \left(\frac{r}{20 r_{\rm g}}\right) \left(\frac{M}{10 M_{\odot}}\right)
 \left(\frac{M+m}{10 M_{\odot}}\right)^{-1/3} \left(\frac{P}{2.62 {\rm d}}\right)^{1/3} {\rm s}.
\label{etime1}
\end{eqnarray}
Here, the orbital velocity was simply evaluated by Kepler's law,
 $v_{\rm orb} = \sqrt{G (M+m)/a}$, where $G$ is the gravitational constant,
 $M$ and $m$ are a black hole and a companion mass, respectively, 
 and $a$ represents the binary separation. 
%
Finally, we adopted $P=2.62$ day for a period of GRO J1655-40 (Bailyn et al. 1995)
 as a sample observational value for eclipsing binary.  
Clearly, a larger black hole mass or longer orbital period
 will increase $\Delta t$. 

The duration time of the total eclipse of the disk is roughly given by
 the following equation; 
\begin{eqnarray}
 t_{\rm eclipse} &\sim & \frac{2 r_*}{v_{\rm orb}} \nonumber  \\
              &=& 10^3 \left(\frac{r_*}{R_{\odot}}\right) \left(\frac{a}{R_{\odot}}\right)^{1/2}
 \left(\frac{M+m}{10 M_{\odot}}\right)^{-1/2}  {\rm s} \nonumber  \\ 
              &=& 4.15 \times 10^3 \left(\frac{r_*}{R_{\odot}}\right) 
 \left(\frac{M+m}{10 M_{\odot}}\right)^{-1/3} \left(\frac{P}{2.62 {\rm d}}\right)^{1/3}  {\rm s}.  
\end{eqnarray} 
Here, $r_*$ is the radius of the companion stars. 

To detect this eclipse time period, $\Delta t$, a high time resolution of 
 at least $\sim 0.01-1$ seconds is required. 
This is not easy to obtain in the optical band;
 however, in the X-ray band, current instrument already
 has sufficiently high time resolution 
 (for example, the $RXTE$ Proportional Counter Array (PCA)
 has a time resolution of milliseconds). 
Observations of eclipsing black hole X-ray binaries
 therefore provide an opportunity to study the physics around black holes.  

\begin{figure}[h]
  \begin{center}
    \FigureFile(55mm,55mm){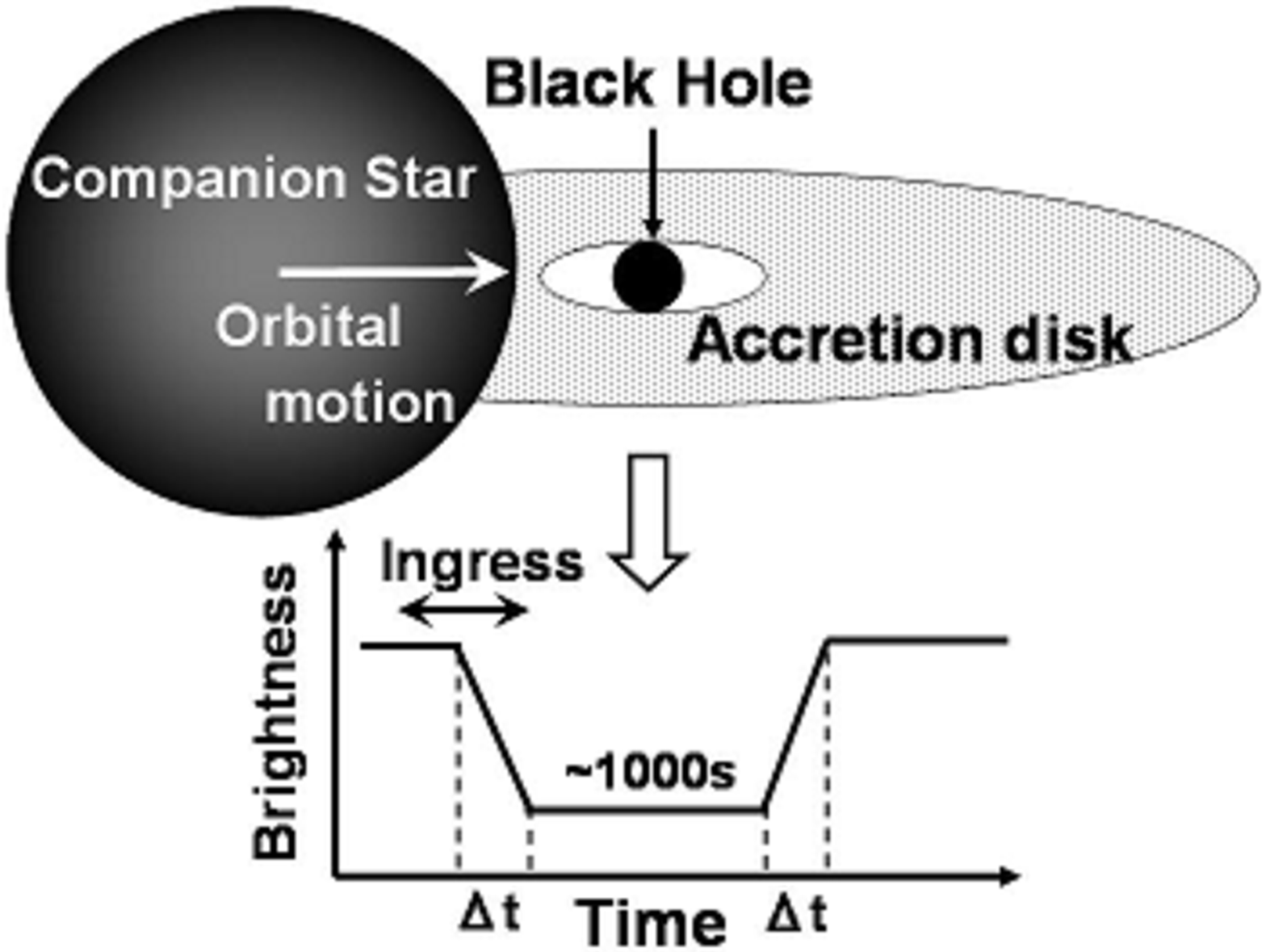} 
    \FigureFile(55mm,55mm){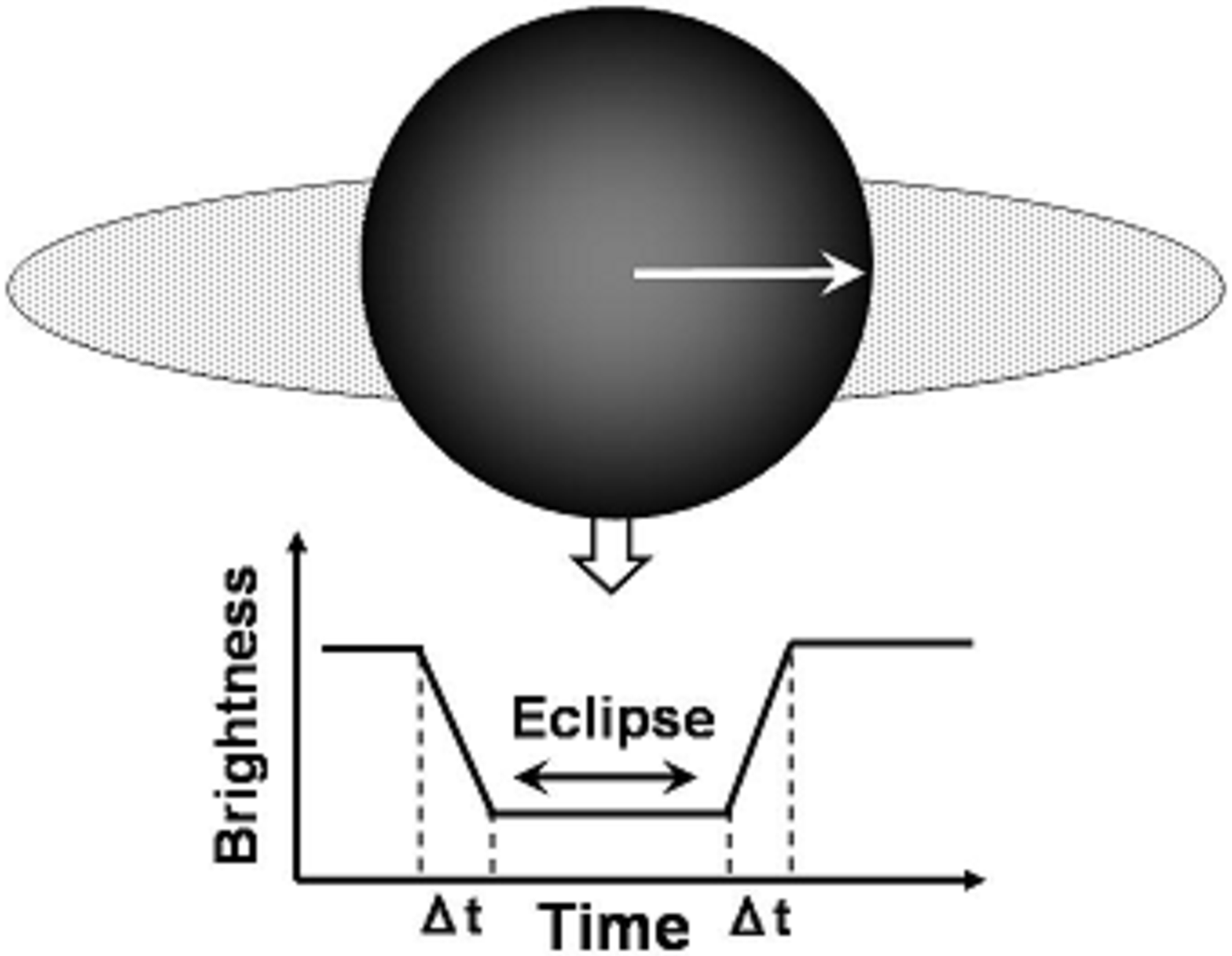} 
    \FigureFile(55mm,55mm){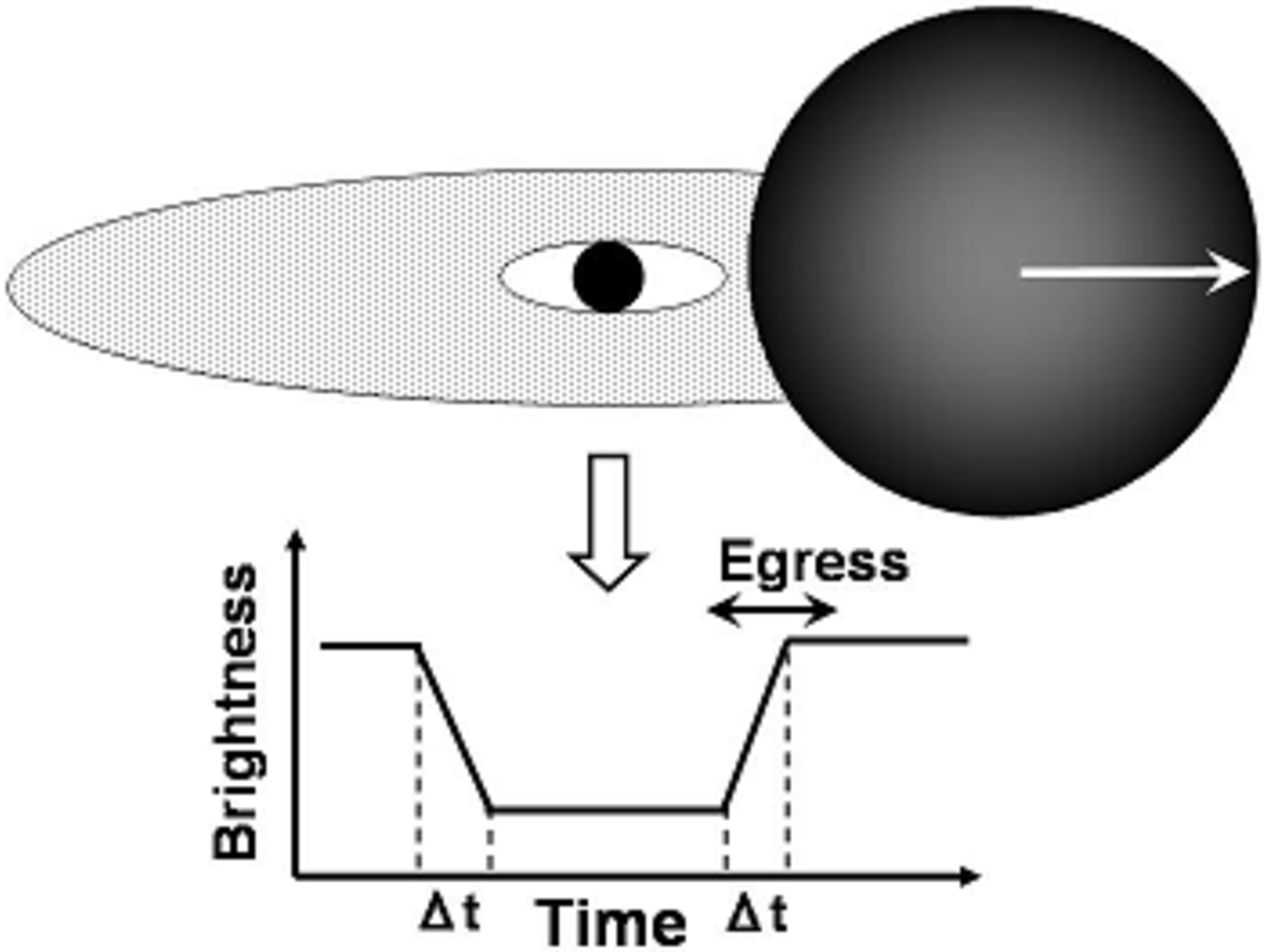}
    \caption{Schematic diagram of our calculation. }
    \label{fig:ponchi}
  \end{center}
\end{figure}
We show a schematic diagram of our calculation in figure \ref{fig:ponchi}. 
Our present study focused on eclipse light curves for a region very close to a black hole ($\lesssim 30~r_{\rm g}$). 
In addition, we implicitly assumed that the binary system has a relatively
 high inclination angle, $i>60^\circ$. 
 
\section{Asymmetric Light-Curve Analysis}

\subsection{Calculation Method}

\begin{figure}[h]
  \begin{center}
    \FigureFile(52mm,52mm){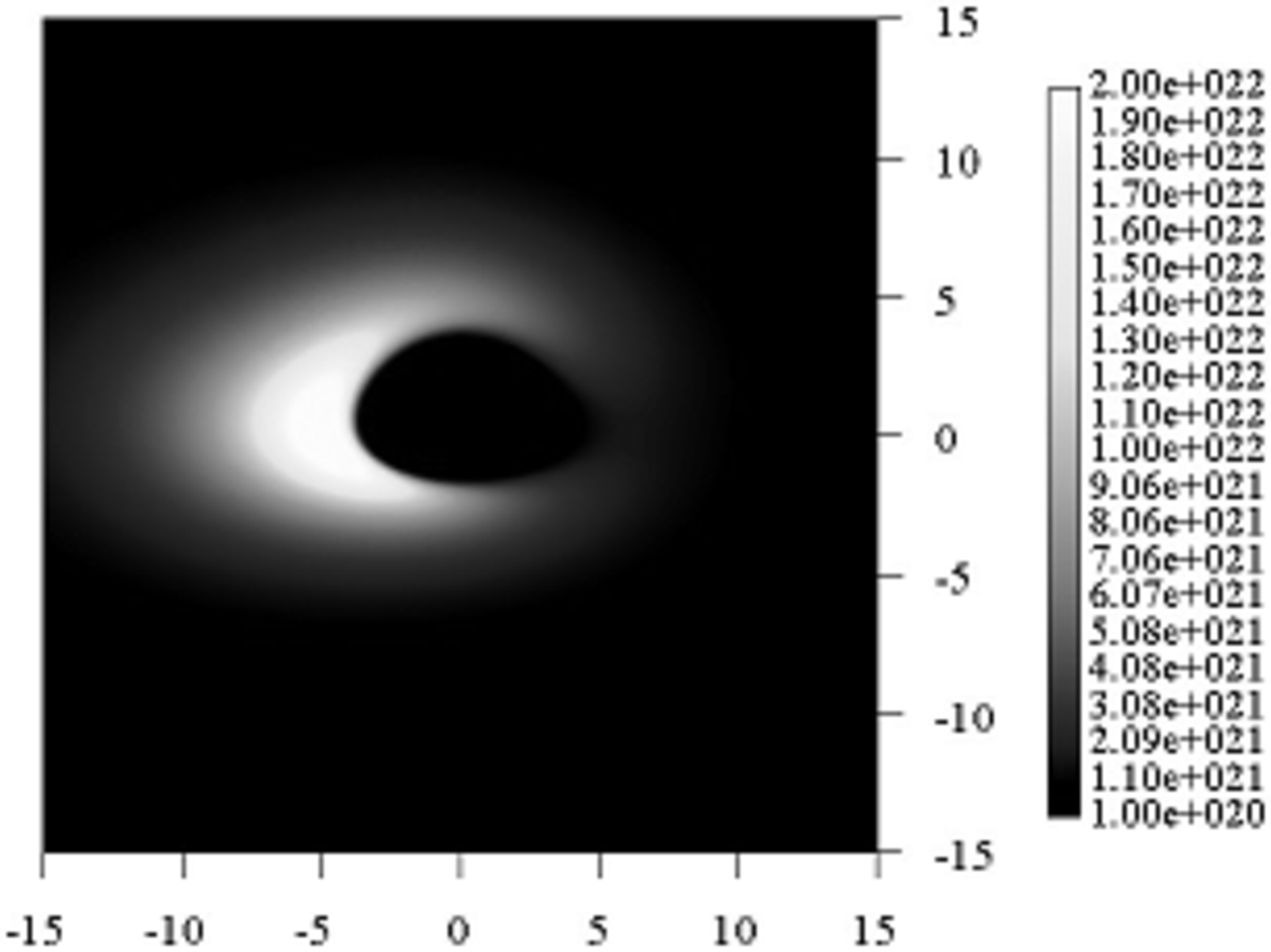}
    \FigureFile(52mm,52mm){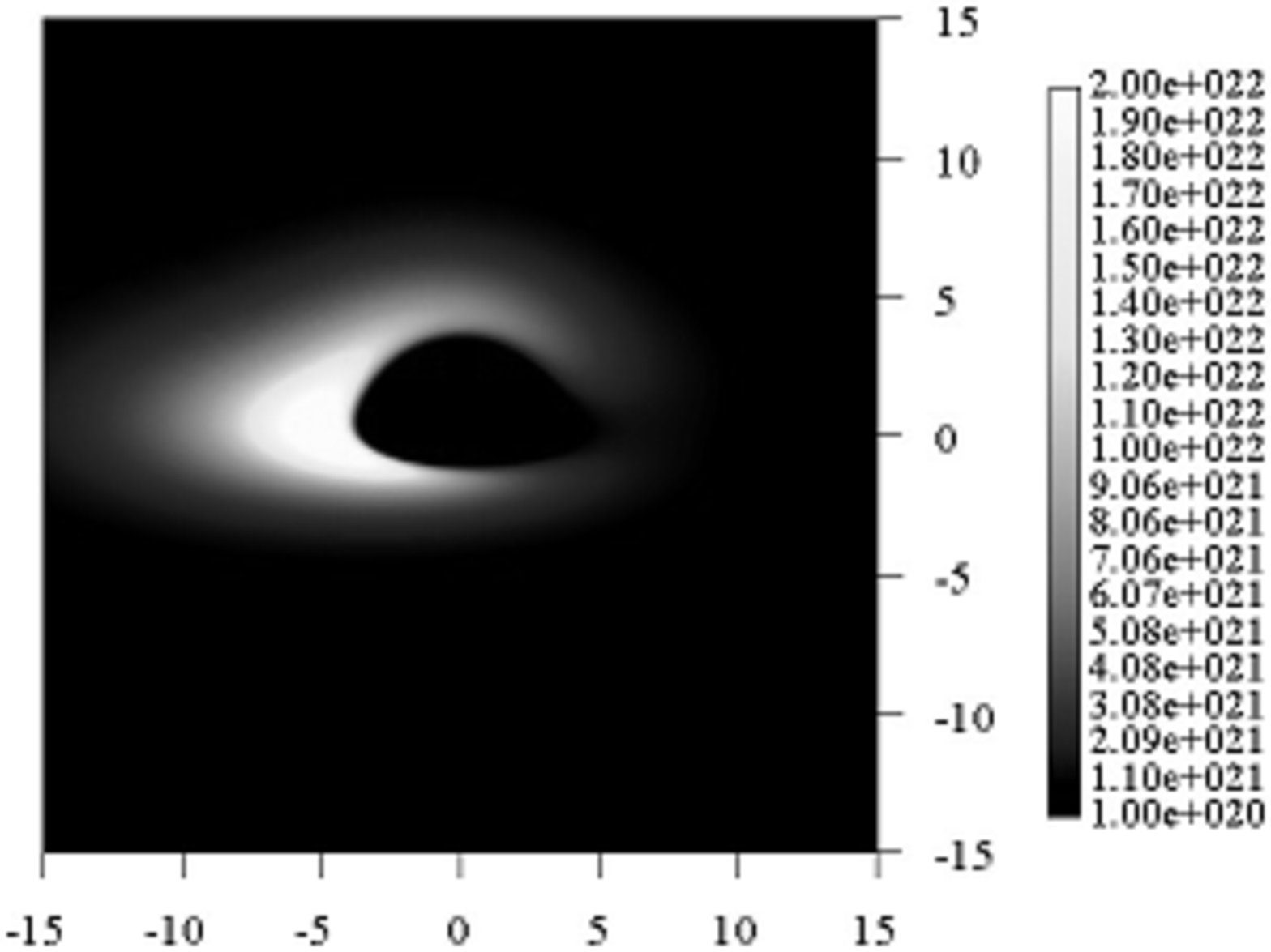} 
    \FigureFile(52mm,52mm){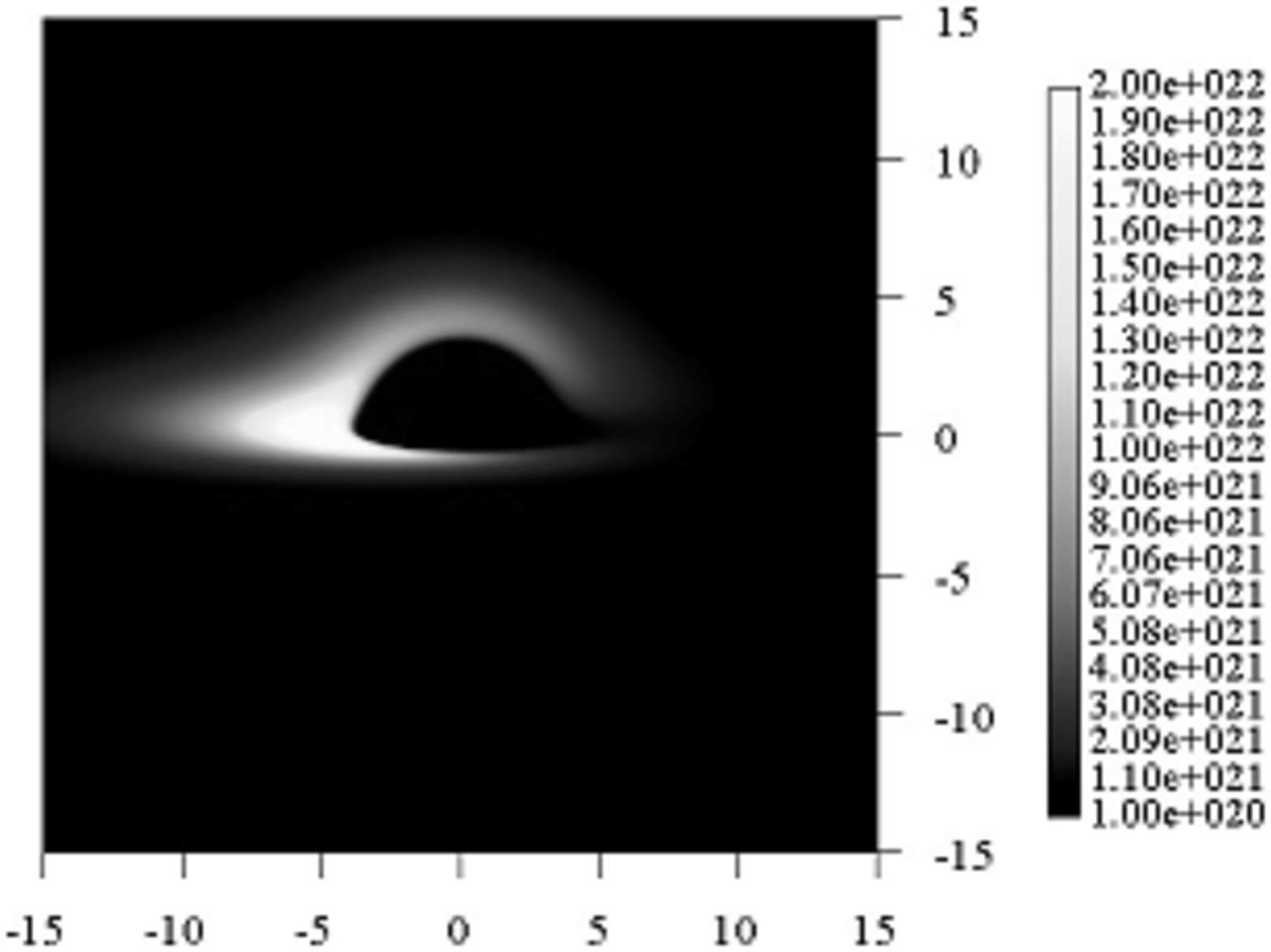} 
    \FigureFile(52mm,52mm){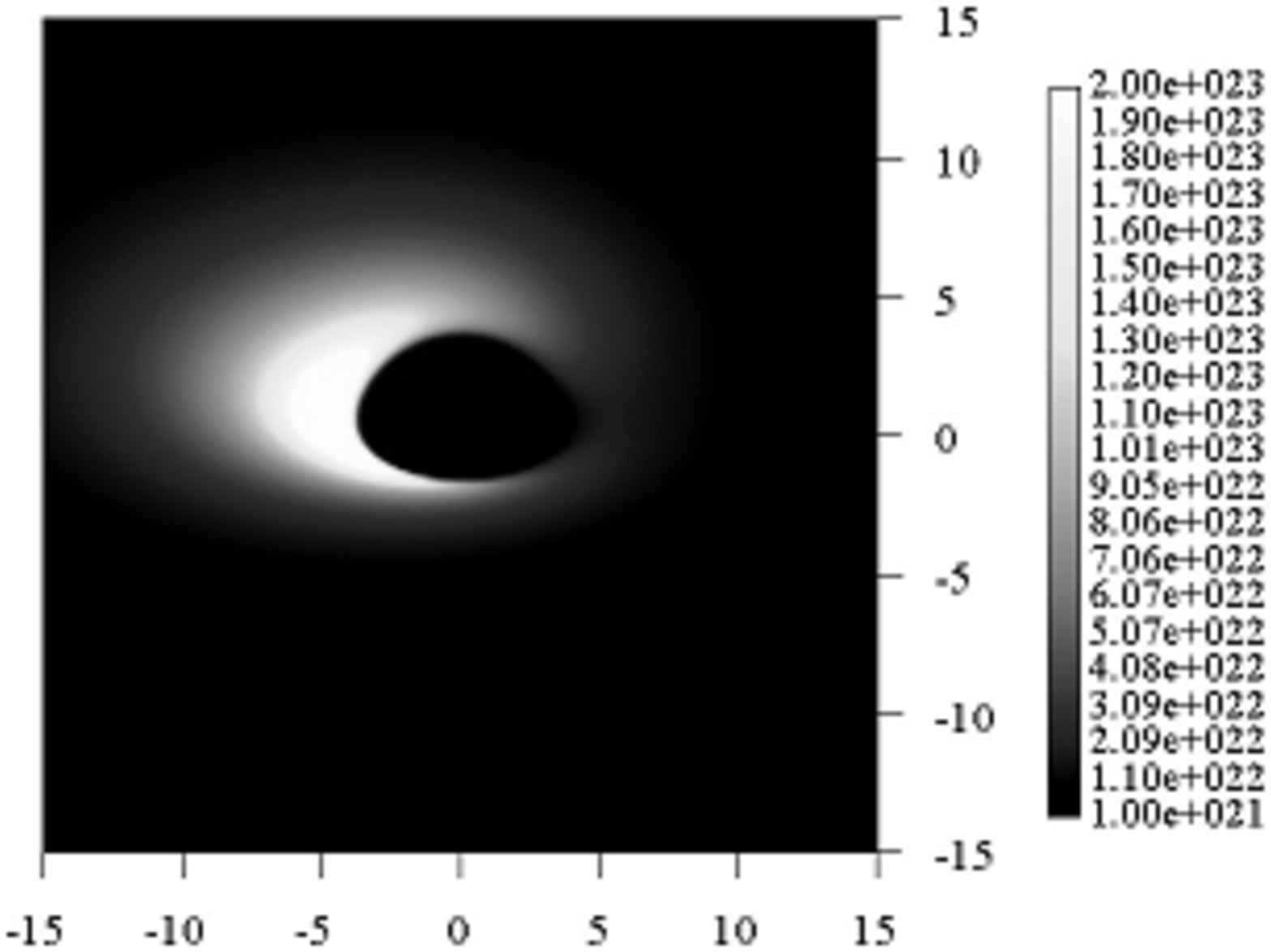} 
    \FigureFile(52mm,52mm){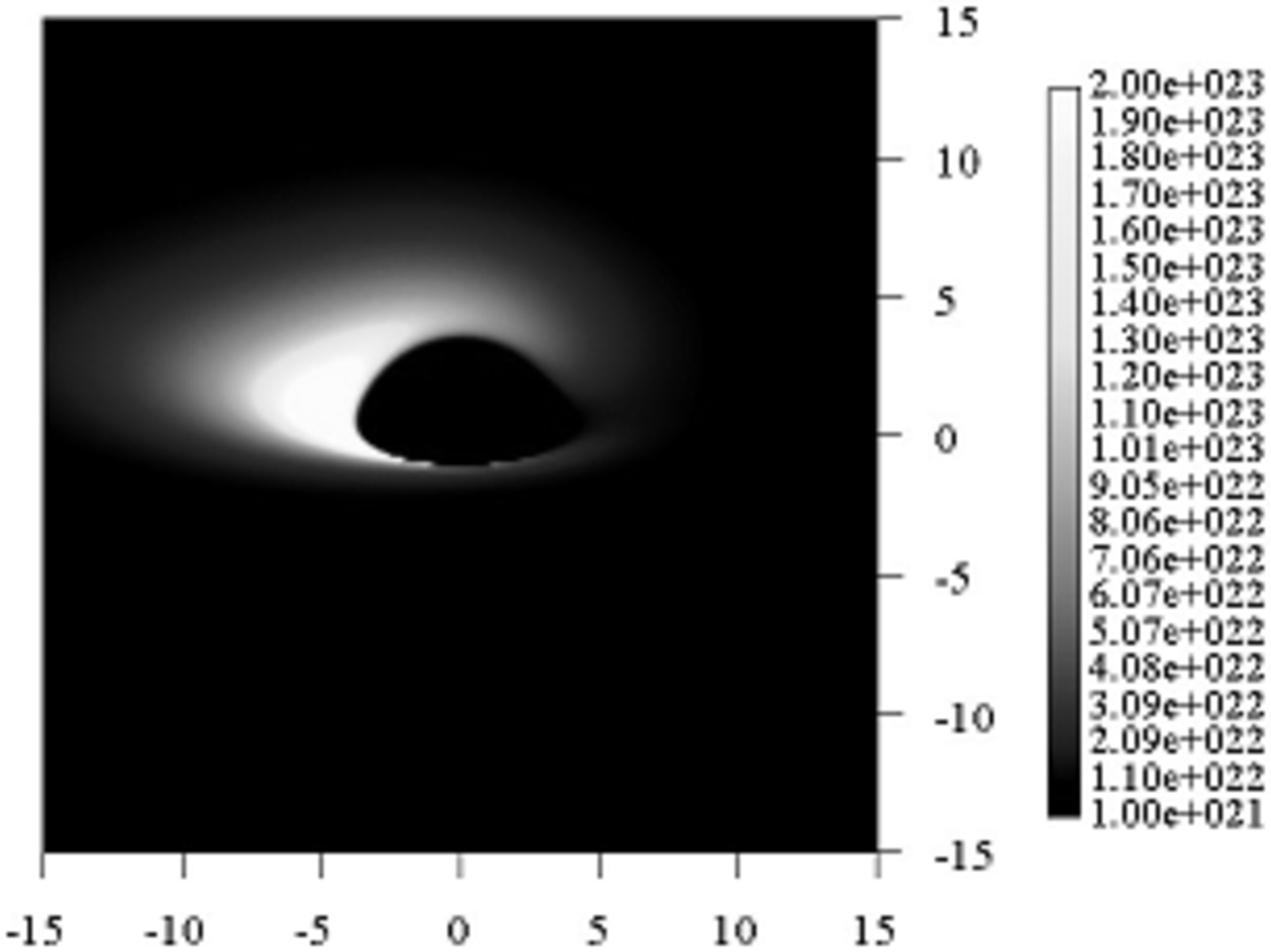} 
    \FigureFile(52mm,52mm){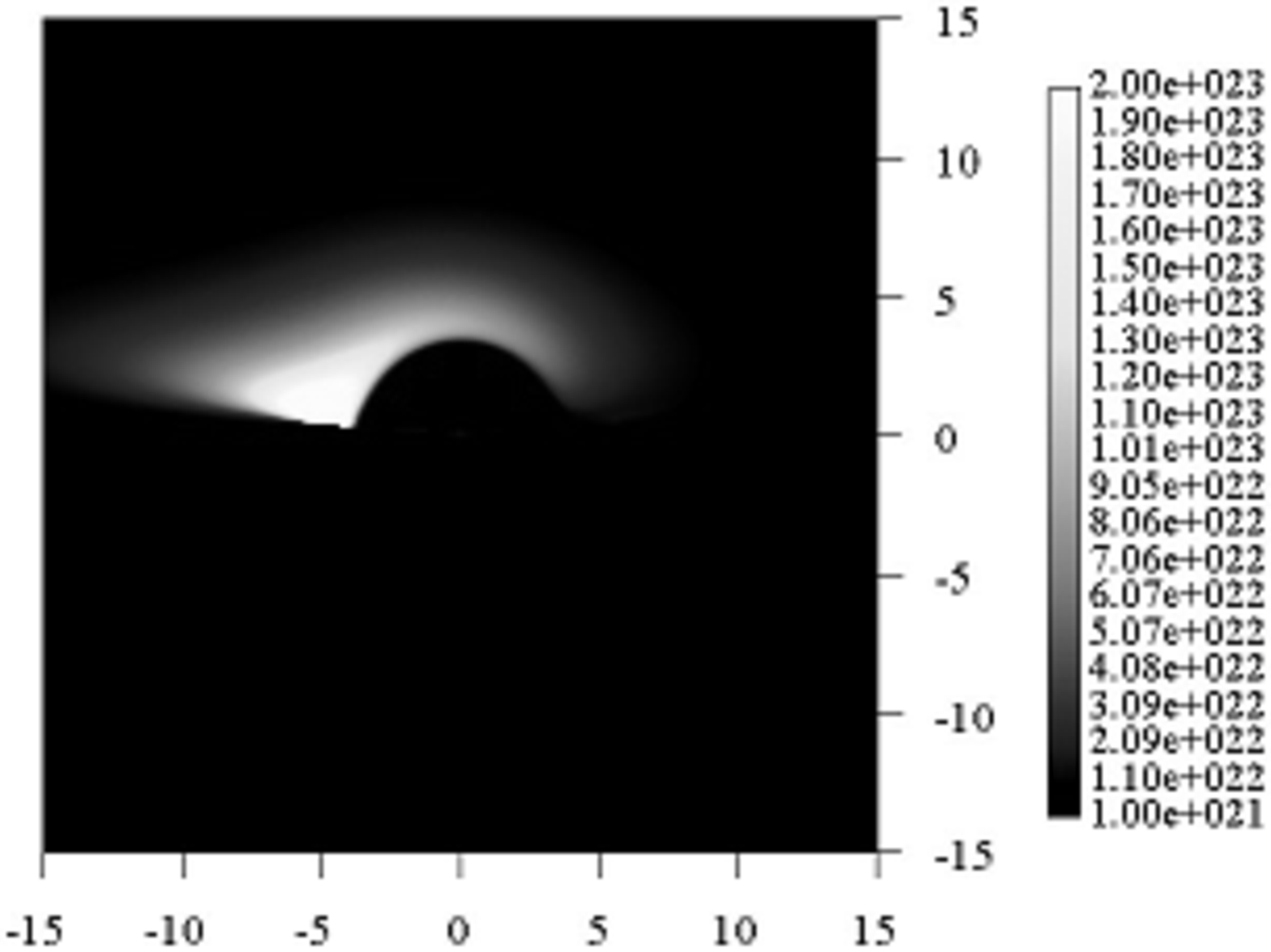} 
    \FigureFile(52mm,52mm){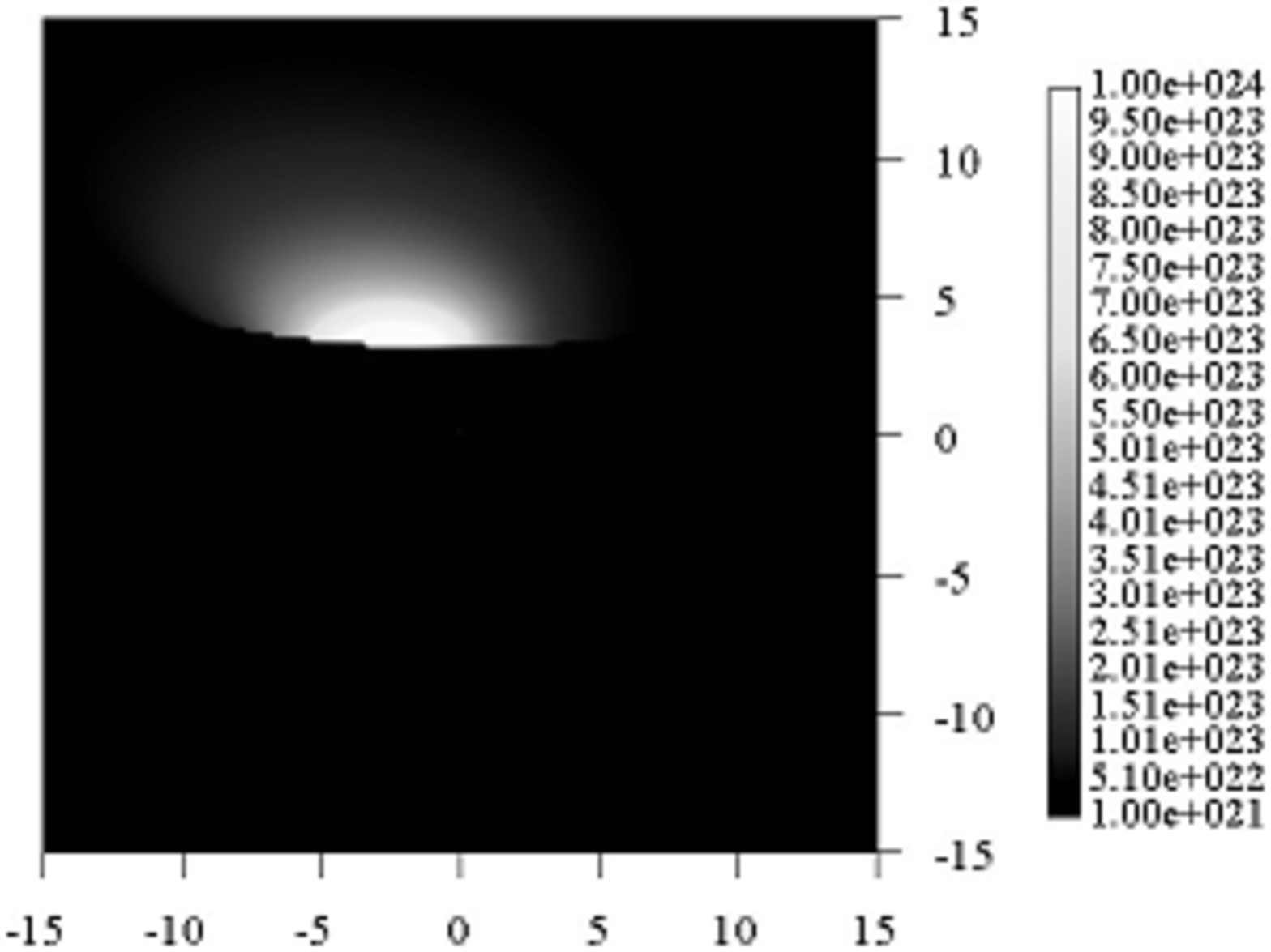}
    \FigureFile(52mm,52mm){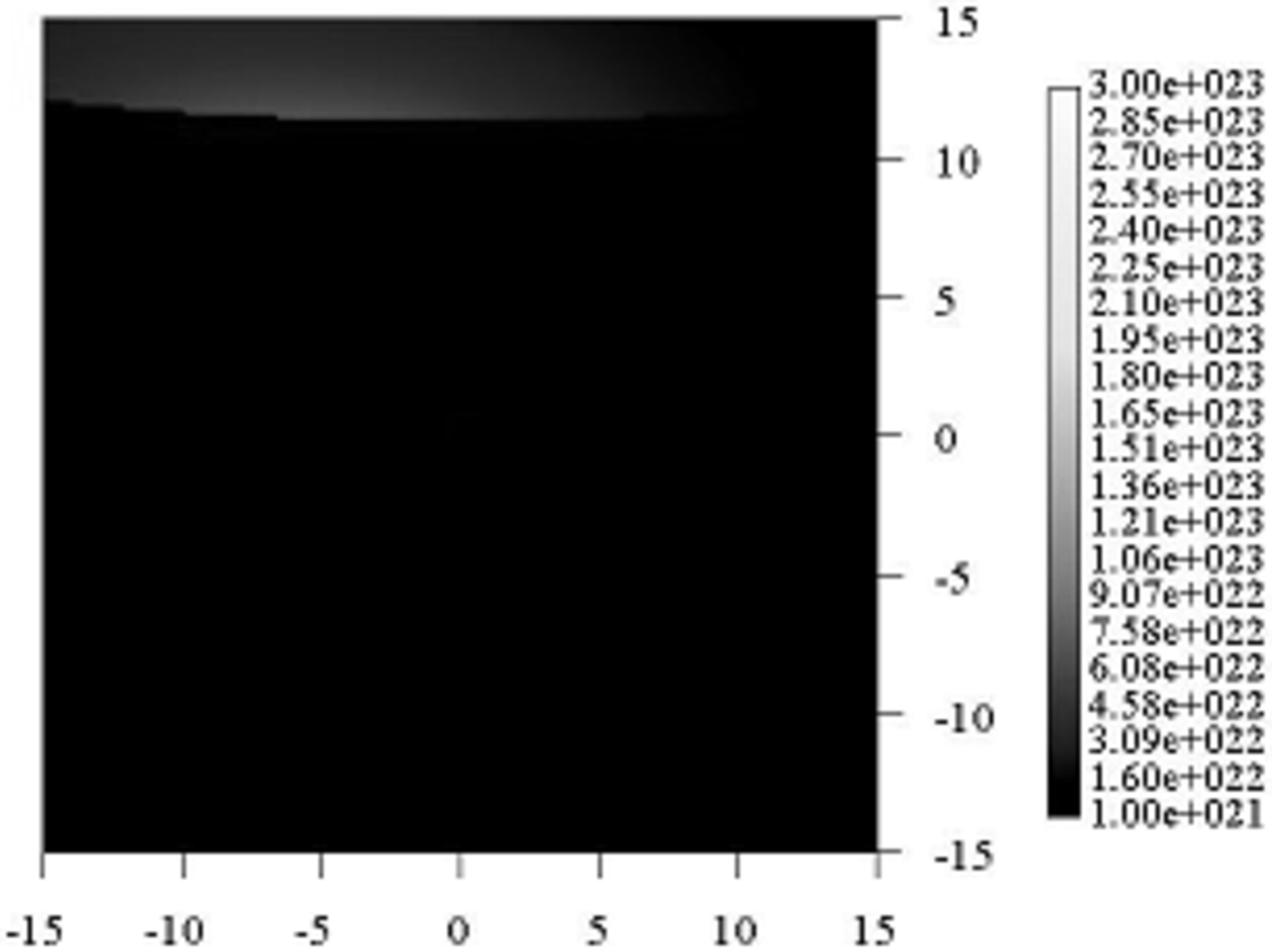} 
    \FigureFile(52mm,52mm){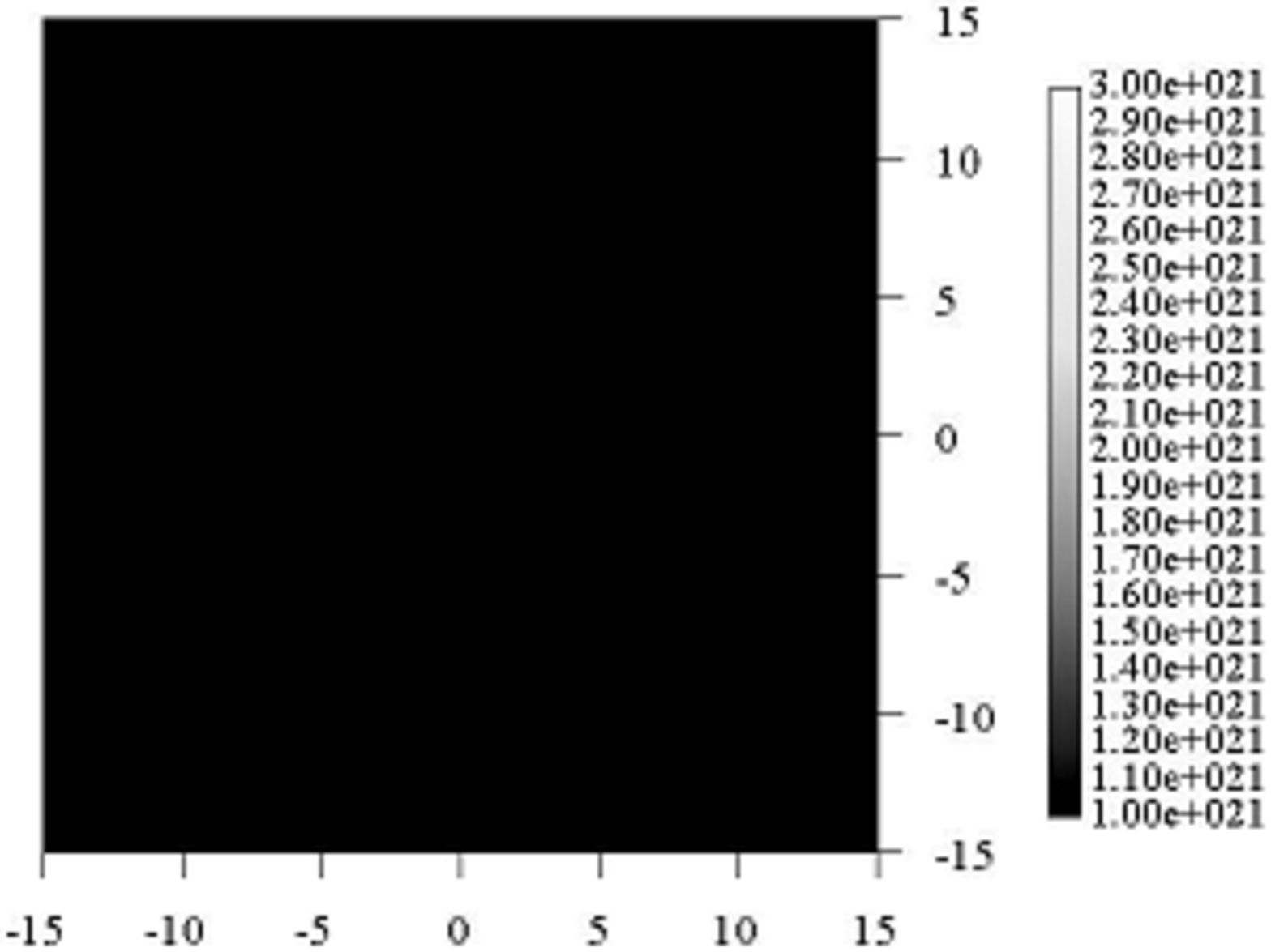} 
    \caption{Bolometric flux distribution with different accretion rates,
 $\dot{m} = \dot{M}/(L_{\rm E}/c^2)$,
 and inclination angles, $i$. 
The right-hand legend of each panel represents
 the bolometric flux level [${\rm erg/cm^2/s}$]. 
 Normalized accretion rates are 1, 10, and 100 from top to bottom. 
 Left column shows $i=60^\circ$, the center shows $i=70^\circ$,
 and the right shows $i=80^\circ$, respectively.
Although the calculation size was $60 r_{\rm g} \times 60 r_{\rm g}$,
 in order to close up the flux distribution
 we plotted $30 r_{\rm g} \times 30 r_{\rm g}$ region. 
Here, $r_{\rm g}$ is the Schwarzschild radius, $r_{\rm g} = 2.95 \times 10^6 (M/10M_\odot)$ cm. 
The bottom right figure (for $i=80^\circ$, $\dot{m}=100$) shows that
 the emission from disk's inner region was completely blocked by the disk's outer region. 
}
    \label{fig:image}
  \end{center}
\end{figure}

Some assumptions were required to calculate a light curve during an eclipse. 
First, we assumed an optically thick relativistic accretion disk
 around a Schwarzschild black hole as a background radiation source. 
We numerically solved a set of hydrodynamical equations
 including the effect of the transonic nature of the flow (Watarai et al. 2000). 
Using the numerical data, we could obtain more realistic temperature profiles,
 velocity field, and disk geometrical thickness unlike simple thin disk solutions.   
For an accretion rate exceeding the Eddington rate, $\dot{M}_{\rm E} = 16 L_{\rm E}/c^2$, 
 the calculated solutions became those of advection-dominated states. 
Such solutions corresponded to so-called ``slim disks'' (Abramowicz et al. 1988),
 and the scale-height of a disk increased as the accretion rate increased. 
When the mass-accretion rate increased,
 the energy generation via viscous heating increased. 
As a result the pressure of the disk became proportional to the mass-accretion rate,
 which gave $H \propto \dot{M}^{1/2}$.  
Hence, the disk became geometrically thick for high mass-accretion rates. 
Using the calculated numerical solutions we computed bolometric flux images
 including the geometrical and relativistic effects (figure \ref{fig:image}). 
Throughout these calculations, we used the normalized accretion rate,
 $\dot{m}=\dot{M}/(L_{\rm E}/c^2)$. 

Second, to obtain an image around a black hole,
 we applied the {\it Ray-Tracing method},
 which is commonly used in astrophysics (see also Fukue, Yokoyama 1988). 
Large numbers of rays were traced from the observer's screen to the black hole, 
 and the rays were calculated along with the null geodesic. 
Accordingly, our calculation automatically included
 the special/general relativistic effects,
 such as the relativistic Doppler effect, photon red shift,
 the light bending effect, etc. 
More details of the calculation method are described in Watarai et al. (2005).

\subsection{Calculated Light Curves}

\begin{figure}[h]
  \begin{center}
    \FigureFile(50mm,50mm){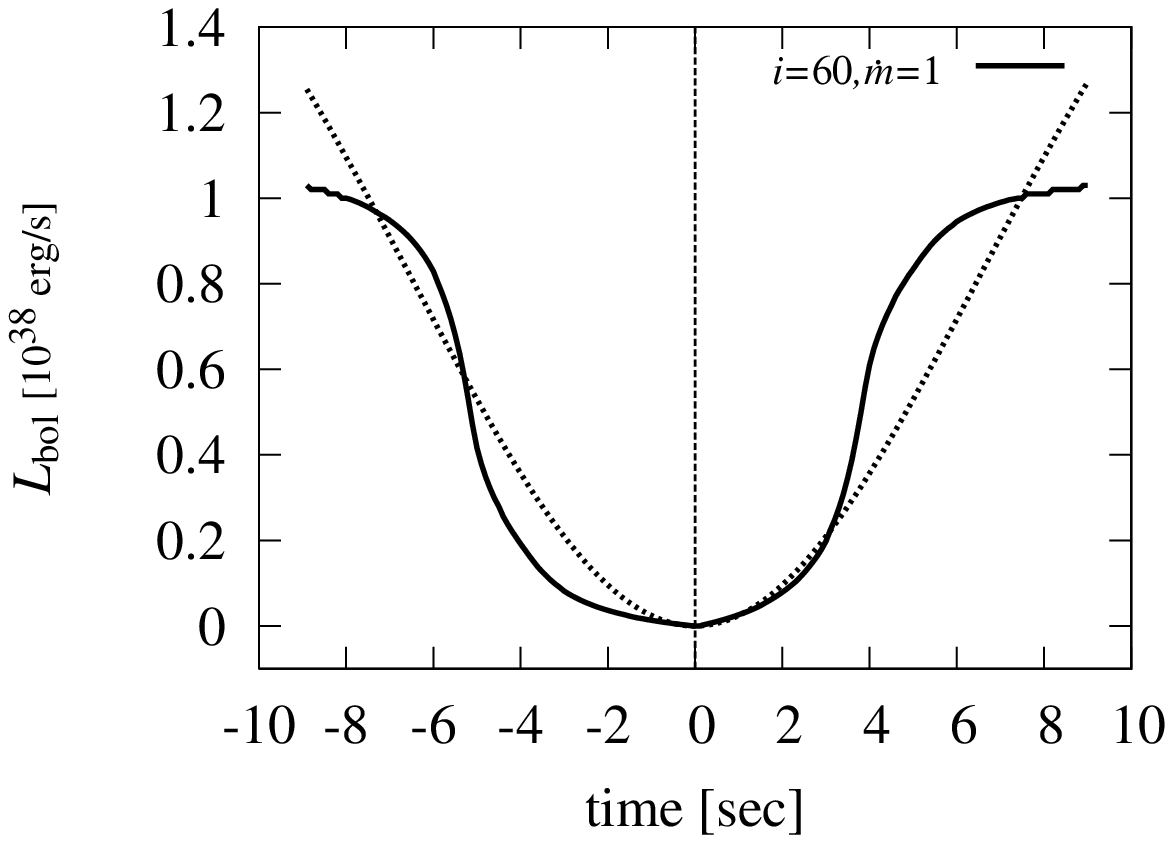}
    \FigureFile(50mm,50mm){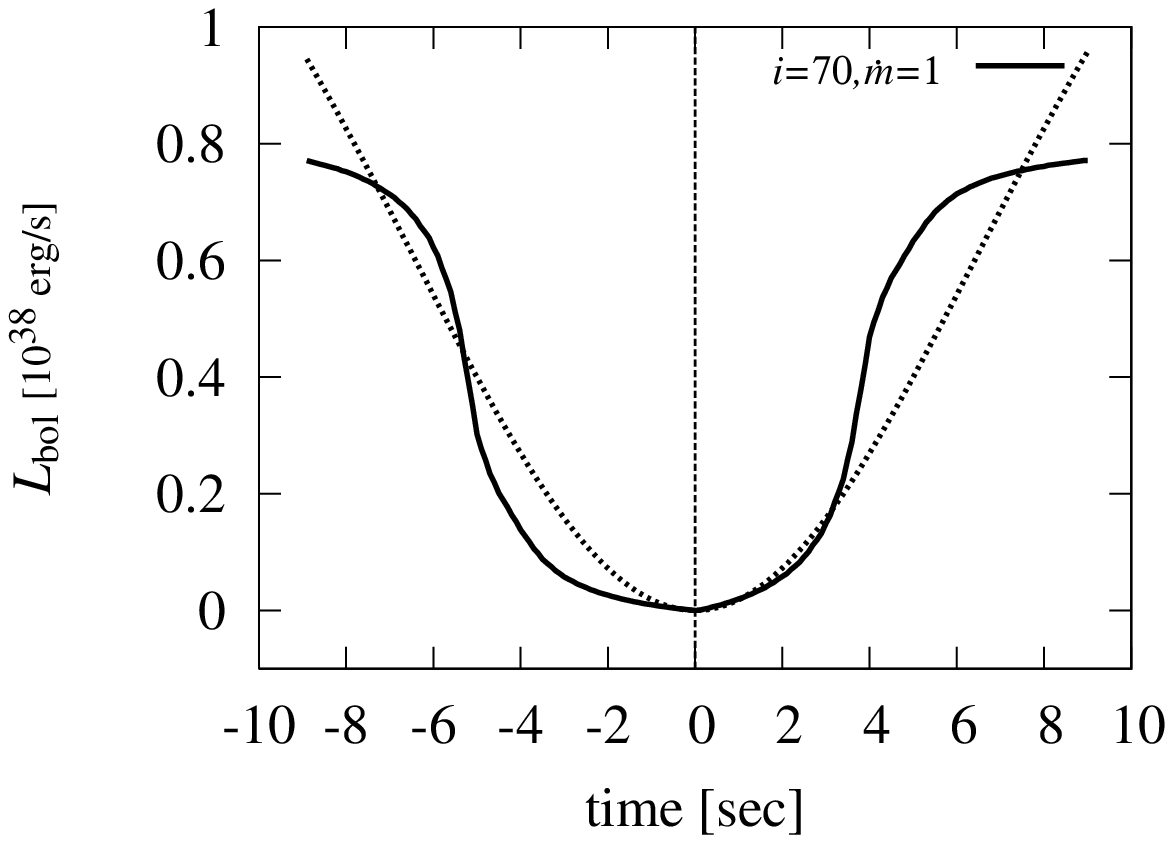}  
    \FigureFile(50mm,50mm){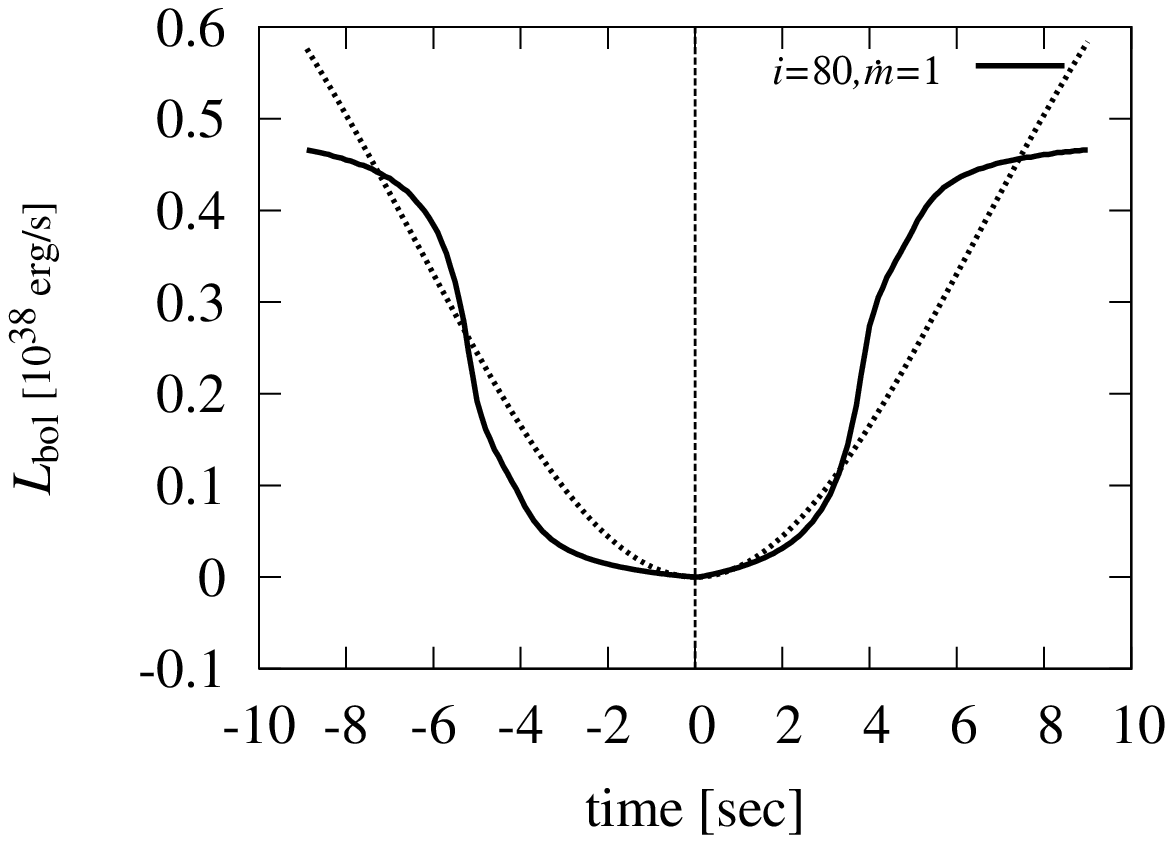}  
    \FigureFile(50mm,50mm){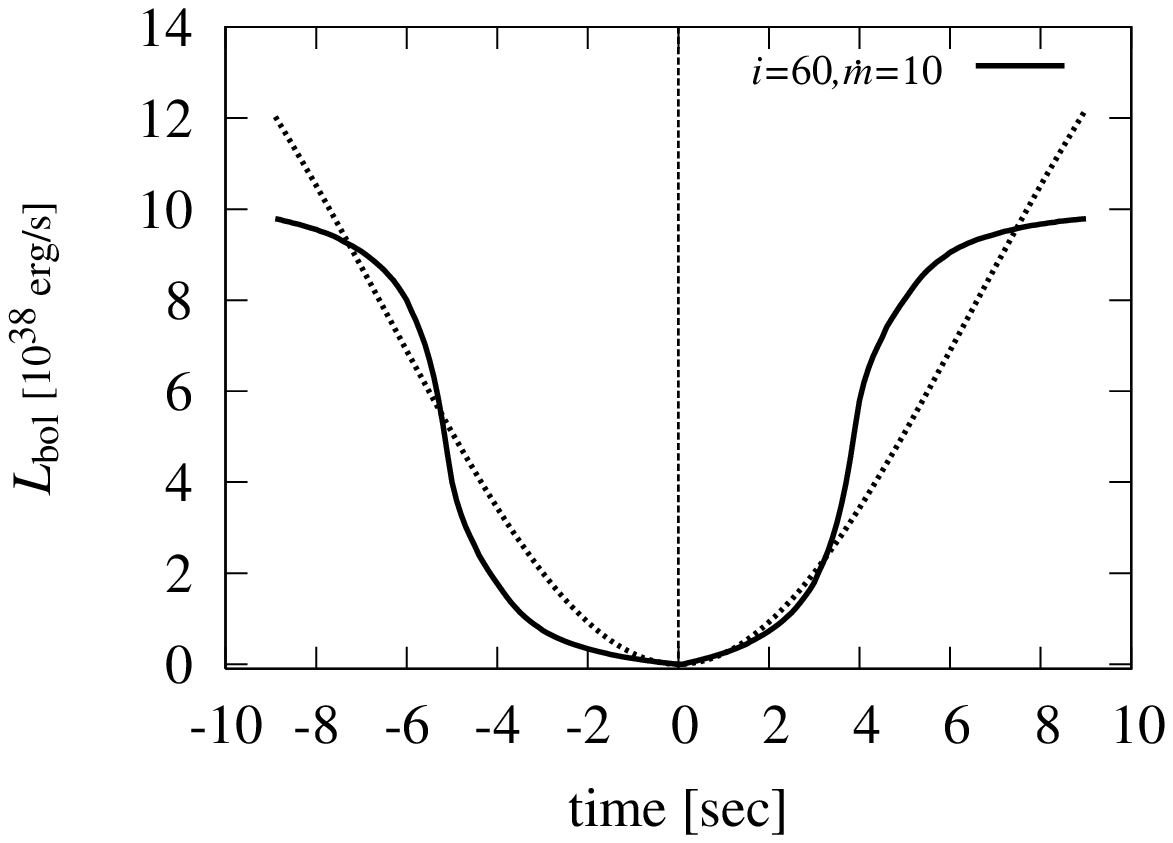}  
    \FigureFile(50mm,50mm){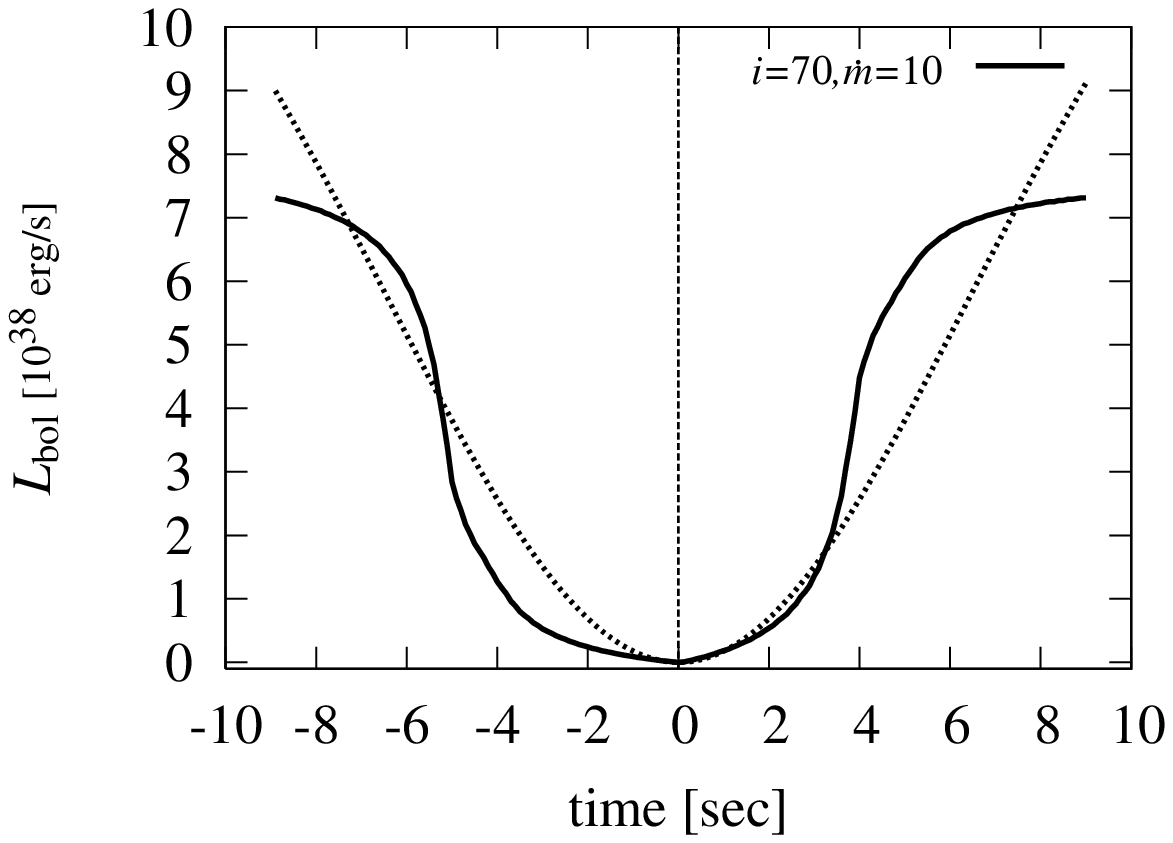}
    \FigureFile(50mm,50mm){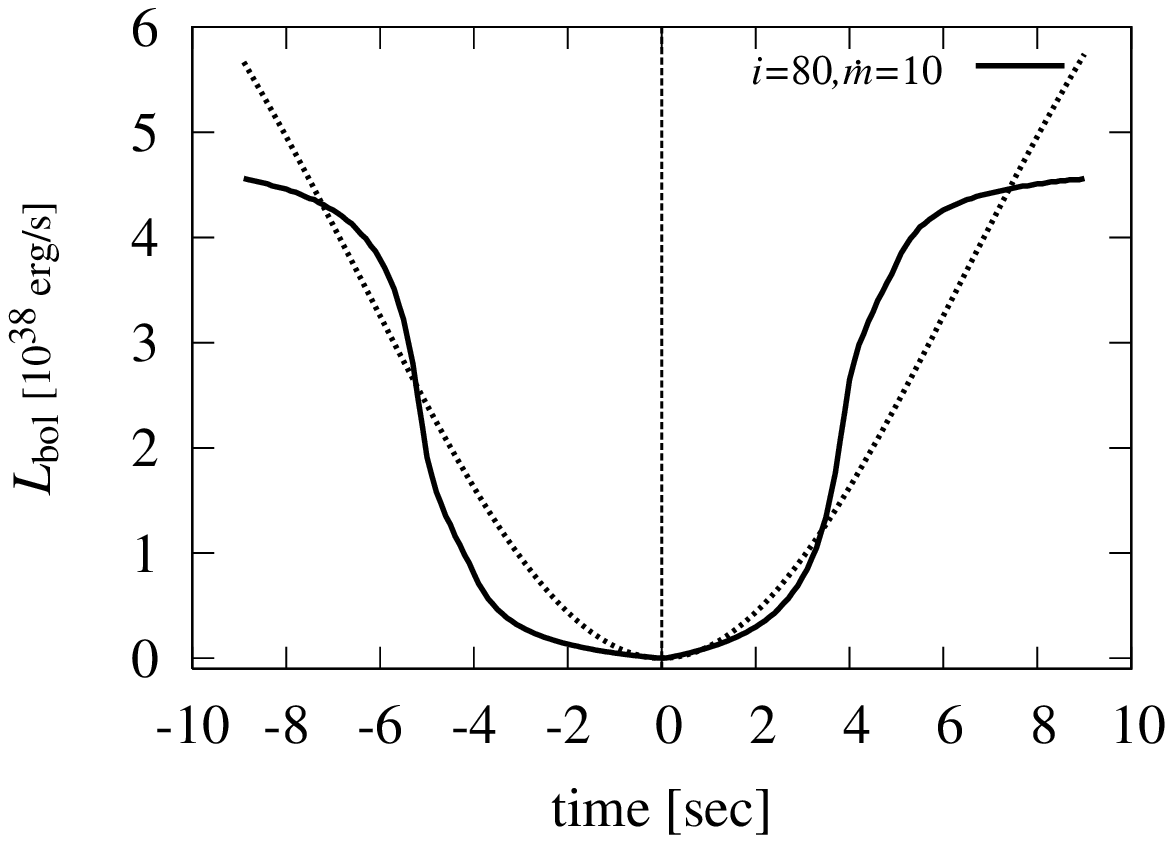}
    \FigureFile(50mm,50mm){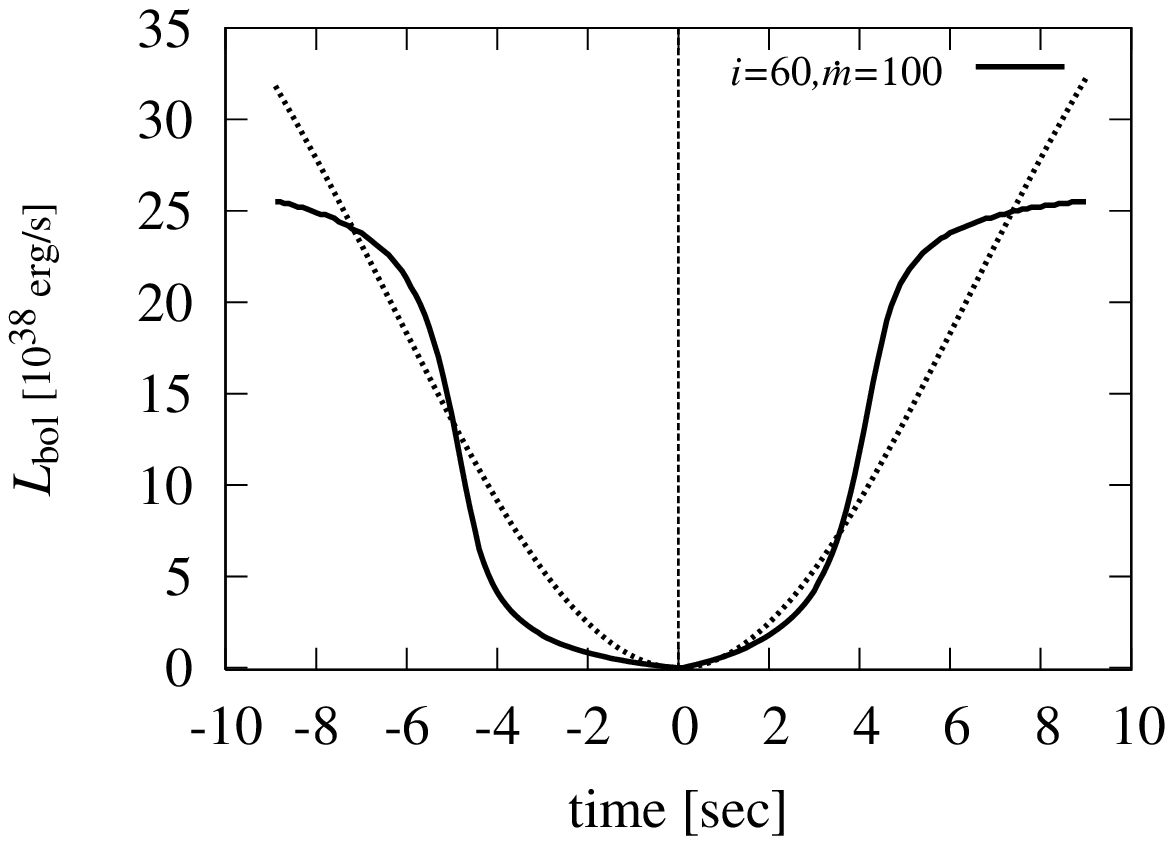} 
    \FigureFile(50mm,50mm){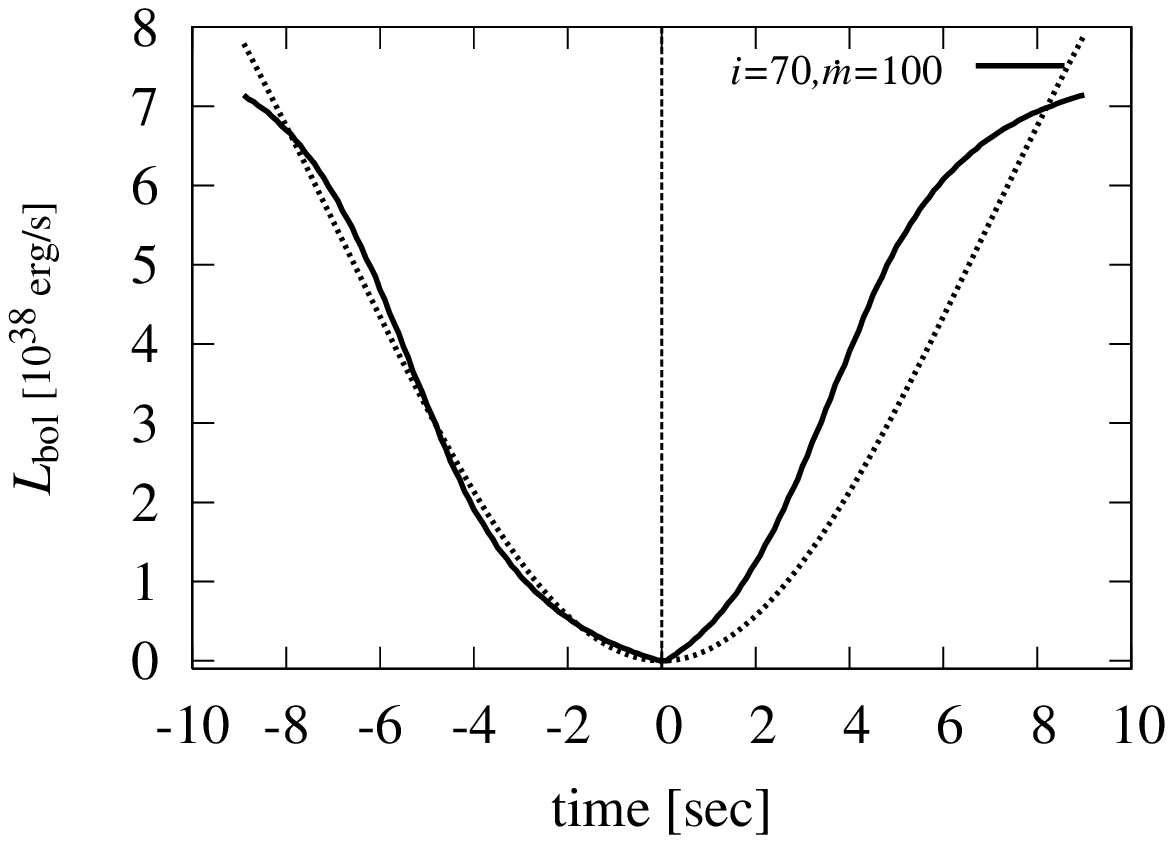}
    \FigureFile(50mm,50mm){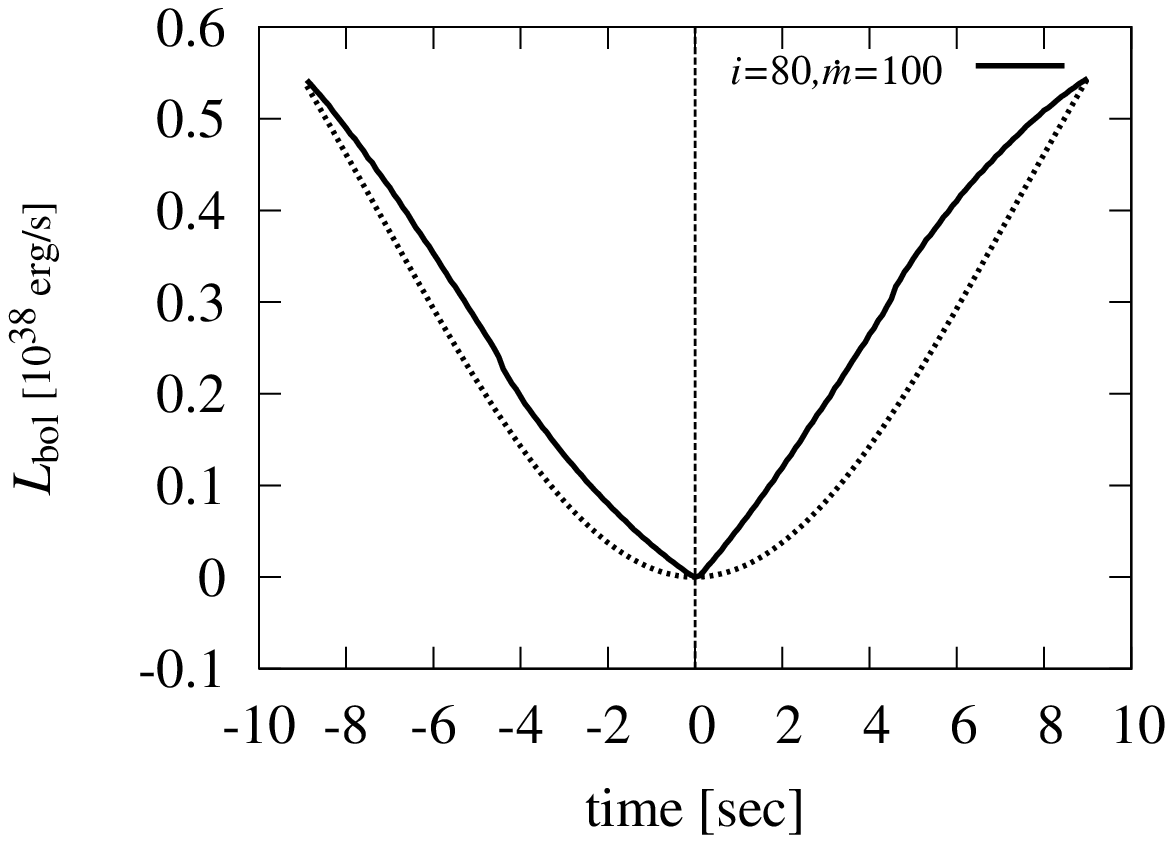}
    \caption{Light curves during before and during (left side of dashed line)
 and after (right side of dashed line) eclipse by a companion star. 
Calculation parameters were the same as those of figure 2; that is,   
 the normalized accretion rates were 1, 10, and 100 from top to bottom,  
 and the inclination angles were $i=60^\circ$, $i=70^\circ$
, and $i=80^\circ$ from left to right. 
Solid lines represent the calculated light curves. 
Dotted lines are inverted from the Gaussian distribution for comparison of axis symmetry. 
Dashed lines showing zero time (time=0 sec) indicate the boundary of complete eclipse. 
We assumed an orbital velocity of 200 ${\rm km~s^{-1}}$. 
}
\label{fig:lc}
  \end{center}
\end{figure}

Figure \ref{fig:lc} shows the light curves at ingress and egress of an eclipse. 
We assumed that the orbital velocity was 200 ${\rm km~s^{-1}}$. 
The total eclipse phase ($\sim 10^3$ s) was removed (see figure \ref{fig:ponchi}),
 and the ingress phase ($\sim 1$ s) was directly joined to the egress phase
 to analyze the eclipse profiles. 
The calculated size was $60 r_{\rm g} \times 60 r_{\rm g}$, 
which is very small relative to the radius of companion star. 
We therefore ignored the curvature of the edge of the companion star
 in calculating the light-curves of the eclipse. 

All of the light curves showed asymmetric profiles (figure \ref{fig:lc}).  
These light-curve features are explained as follows.  
At ingress, initially the brighter part was blocked 
 so that the light curve first decreased rapidly and then more gradually. 
On the other hand, at egress, the brighter part appeared first,
 and thus the light curve increased rapidly and then more gradually. 
However, there are several differences in these asymmetric light curves.
In particular, there is a dependence on inclination angle, $i$: 
 the asymmetry strengthens as $i$ increases. 
This is due to the emission from high-$i$ accretion disks
 becoming amplified by relativistic Doppler beaming. 

As the accretion rate increased, the shape of the light curve became symmetric 
because the emission from the disk's inner region suffered from the self-occultation
 of the disks outer rim. 
Therefore, the light curves retained a normal distribution
 due to the absence of an asymmetric emitting region. 
The degree of asymmetry of these light curves
 will be discussed in the next subsection. 

\subsection{Skewness and Kurtosis}

\begin{figure}[h]
    \FigureFile(80mm,80mm){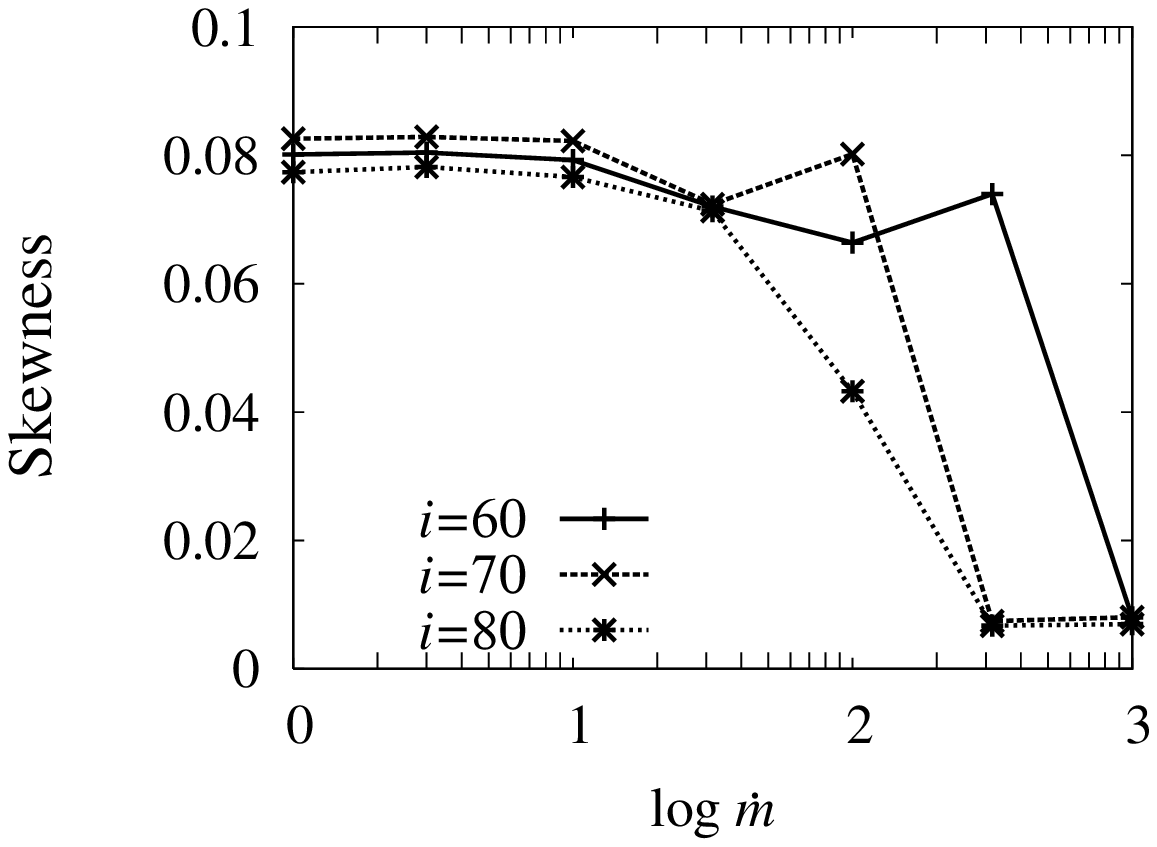} 
    \FigureFile(80mm,80mm){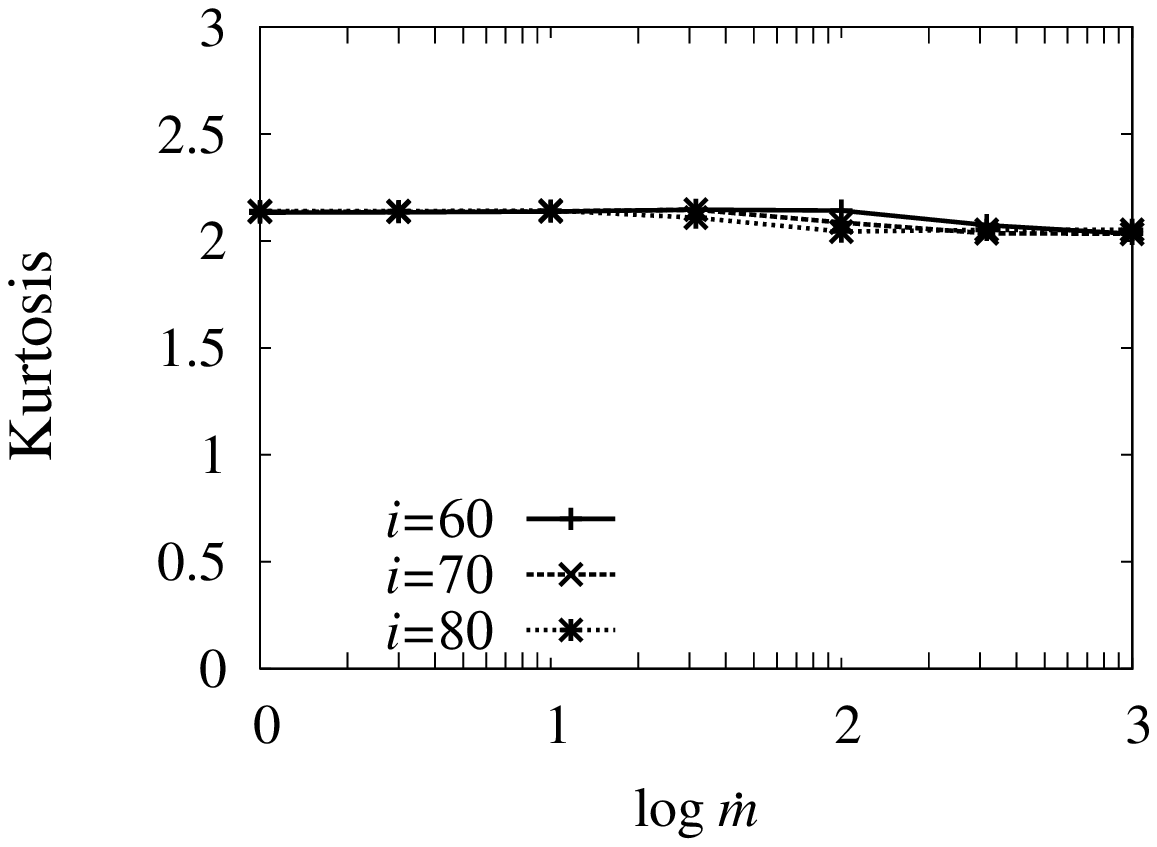} 
    \caption{The skewness and kurtosis for different accretion rates. 
Solid lines represent $i=60^\circ$ case, dashed lines represent $i=70^\circ$,
 and dotted lines represent $i=80^\circ$ case. }
\label{fig:toukei}
\end{figure}

Asymmetric light curves will be observed when the object is sufficiently bright
 and when a companion star crosses in front of a black hole during an eclipse. 
Here, we seek a simplified observational indicator that expresses
 the degree of asymmetry. 
We introduce the statistical quantities of skewness, $S$,
 and kurtosis, $K$, as indicators of light-curve asymmetry. 
The skewness and kurtosis represent the deviation
 of the observational data from the Gaussian (normal) distribution. 
The skewness measures the asymmetry of the distribution. 
The kurtosis measures how sharply peaked a distribution is relative to its width. 

First, we inverted the light-curve data to obtain $f_i$. 
Then, we used the modified data, $f_i$ for skewness-kurtosis analysis. 
The statistical quantities were defined as follows:  
\begin{eqnarray}
N &=&  \sum_{i=1}^n f_i,  \\
\bar{x} &=&  \frac{1}{N} \sum_{i=1}^n x_i f_i, \\
\sigma &=&  \frac{1}{N} \sum_{i=1}^n (x_i - \bar{x})^2 f_i, \\ 
S &=&  \frac{1}{N \sigma^3} \sum_{i=1}^n (x_i - \bar{x})^3 f_i,  \\
K &=&  \frac{1}{N \sigma^4} \sum_{i=1}^n (x_i - \bar{x})^4 f_i. 
\end{eqnarray}
Here, $N$ is the number of the data points, 
  $x_i$ is the time of each mesh (sample numbers of an observation),
  $f_i$ is the frequency of the data, $\bar{x}$ is an average,
 and $\sigma$ is the standard deviation. 
In our analysis, 
 $x_i$ and $\bar{x}$ are the time of our observation and its average
 (which is the same as half time of the total eclipse time), 
and $f_i$ corresponds to the amplitude of observed/calculated flux.  
When the data agree with the normal distribution,
 the distribution is symmetric, i.e., $S$ equals to zero. 
A positively skewed distribution ($S > 0$) has a longer tail to the right;
 whereas, a negatively skewed distribution ($S < 0 $)
 has a longer tail to the left. 
 
The calculated skewness and kurtosis are plotted in figure \ref{fig:toukei}. 
A large value of skewness indicates a high degree of asymmetry. 
The skewness for a small accretion rate, $\dot{m}$=1--10,
 remained constant value even for different inclination angles. 
This is because in this situation we could clearly observe
 an asymmetric brightness distribution around the black hole.  
In this regime, the ratio between blue shift and red shift
 did not alter significantly, so that
 the inclination angle did not affect the skewness value. 
If this characteristic skewness values, $S \sim 0.08$, were detected,
 then it would be an observational effects associated with the black hole. 
On the other hand, the kurtosis value did not change significantly
 with accretion rate, i.e., the variation amplitude was very small. 
This means that the sharpness of the profiles is almost the same over
 a wide range of accretion rates. 
Consequently, the kurtosis is not a useful indicator of relativistic effects.

Let us consider the causes of the change of skewness variation in more detail. 
When the accretion rate of the disk is less than the sub-critical rate
 ($\dot{m} \lesssim 10$), 
the rotational velocity is much higher than the radial velocity,
 $v_{\varphi} \gg v_r$. 
Therefore, Doppler boosting via the rotational velocity 
 component is dominant, and the asymmetry of the light curves becomes large. 
As a result, the skewness attains relatively large values $\sim 0.08$. 
On the other hand, when the accretion rate increases, 
 the radial velocity of the flow approaches to the same order as 
 the rotational velocity. 
However, Doppler boosting via the radial velocity component 
does not contribute to the asymmetry of the light curves;
 it mainly amplifies the observed flux. 
For supercritical accretion flows,
 the rotational velocity decreases while the radial velocity increases, 
 since the angular momentum loss via viscosity becomes effective. 
As a result, the flux distribution on the X-Y plane approaches to be symmetry. 
Accordingly, the degree of light-curve asymmetry decreases
 for supercritical accretion flows, and the skewness approaches zero. 
 
A small increase in skewness can be seen
 for ($\dot{m}$, $i$) = (100, $70^\circ$) and (316, $60^\circ$)
 in figure \ref{fig:toukei}. 
These parameter sets, that is,
 ($\dot{m}$, $i$) = (100, $70^\circ$) and (316, $60^\circ$),
 are marginal values at which 
 the self-occultation of the disk begins to be effective. 
In these parameter sets, the radiation area on the side near to the observer
 was concealed by the disk outer rim,
 but the radiation area on the opposite side was not concealed. 
As a result, the small increase in skewness was observed.

Meanwhile, as the all mass-accretion rate increases,
 all light curves become symmetric
The is because the inner region of the disk was entirely unobservable 
 for high inclination angle.  
In other words,  due to the self-occultation of the disk, 
 we could not observe the relativistic Doppler beaming
 which causes the asymmetric flux distribution,
 and so the skewness approaches zero. 
This geometrical effect occurs only in the cases of
 high mass-accretion rates ($\dot{m} \gtrsim 32$). 
Hence, the degree of the skewness can be an indicator of accretion rate.

\section{Discussion}

\subsection{Is this Diagnosis Useful for the Verification of Relativistic Effects?}

In our study, we adopted a set of statistical quantities, i.e., skewness and kurtosis,
 as indicators of relativistic effects.  
The skewness $S \sim 0.08$ 
 did not alter significantly for small accretion rates and relatively high inclination angles
 ($i=60^\circ - 80^\circ$). 
This means that it is possible to observe relativistic effects,
 i.e., an asymmetric brightness distribution. 

For high-luminosity objects, 
 it will be possible to observe such asymmetric features in the X-ray band. 
For example, the {\it RXTE} PCA has sufficient time resolution to observe
 the eclipse phenomenon, which changes the flux on a timescale of 0.01 seconds. 
It would be difficult to observe this asymmetry in the optical band
 even if optical telescope detectors had sufficient time resolution. 
Although the temperature around the inner-disk region is higher than that of the outer disk,
 the surface area of the inner-disk region is much smaller than that of the outer region. 
Therefore, the dominant emission in the optical band comes from the outer region. 
With an observational instrument having a spatial resolution
 of less than $\sim 100~r_{\rm g}$, in principle, 
 the light-curve asymmetry could be detected even in the optical band. 

The emission from the vicinity of the event horizon contains the information 
about the physical parameters of a black hole. 
In particular, the shape and the position of a black hole can be deformed
 by its rotation and its charge (Takahashi 2004, 2005). 
In principle, light curves of the occultation of the black hole
 may also contain the information about the physical parameters of the black hole. 
Using the light-curve analysis described here, it may be possible to measure
 the black-hole spin using light-curve analysis (Takahashi et al. 2005).  
This is because a rotating black hole transfers angular momentum to
 the accreting gas via the frame-dragging effect, 
 and so the marginal stable circular orbit decreases to less than 3 $r_{\rm g}$. 
Hence, light curves for rotating black holes have a gentler slope at ingress or egress,
 compared to non-rotating black holes.
Verification of the expected difference in the light-curve slopes of
 rotating and non-rotating black holes is an intriguing topic for future study.

\subsection{Finding an Eclipsing Black Hole}

A major problem remains regarding the light-curve analysis we described here: 
 finding suitable observational targets. 
Can we identify a black hole that is covered by its companion star? 
How many eclipsing black hole X-ray binaries exist? 
To date, these questions remain unanswered. 

To confirm the predicted asymmetric features of light curves, 
 it is necessary to find an X-ray eclipse in a black hole candidate system. 
Recently, an eclipsing black hole X-ray binary was observed in M33 X-7 (Pietsch et al. 2004). 
This object does not have pulsations, which are commonly seen in X-ray pulsars, 
 and the mass of the compact object derived from orbital parameters
 exceeds 2.1--3.0 $M_\odot$ for $i=90^\circ$. 
For these reasons the compact object is suspected to be a black hole. 
If a large number of X-ray eclipse black hole candidates are
 discovered in our galaxy, as well as in external galaxies in the near future,
 it will be possible to extend the search for stellar-mass black holes
 by the use of the light-curve analysis presented here. 

Table 1 summarizes the eclipsing black-hole candidates identified to date. 
The ingress or egress times ($\Delta t$) for these objects are observable with 
 the time-resolution of current X-ray telescopes. 
In particular, $\Delta t$ in GRS 1915+105 is about $10$ seconds,
 which is sufficiently large to be observable not only with X-ray telescopes,
 but also in the optical telescope. 
To date, no X-ray eclipse has been observed in GRS 1915+105.
However, the limited observations of this object do not allow us
 to decide whether it is an eclipsing system.  
We therefore look forward to long-term observation that
 covers the whole orbital period of the binary systems, 
 or to observations by next-generation X-ray telescope missions. 

\subsection{Observational Constraints for Accretion Disks}

When light-curve asymmetry is undetected,
 what can we infer about the binary system?
One possibility is self-occultation of the disk. 
As in figure \ref{fig:image}, when the mass-accretion rate is high ($\dot{m}=100$)
 the asymmetric intensity pattern does not appear
 because of the self-occultation effect. 
If we assume this scenario, we can interpret an object
 having a symmetric light curve at eclipse as a super-critical acccretor. 
For example, SS433 is thought to be undergoing
 supercritical accretion, $\dot{m}\gtrsim 32$,
 (Cherepashchuk et al. 1982; Gies et al. 2002; Revnivtsev et al. 2004). 
We therefore predict that the skewness of the light-curve for SS433 will be close to zero. 
Using the result in figure \ref{fig:toukei}
 we can constrain the minimum value of the accretion rate. 
Unfortunately, the X-ray emission from SS433 contains 
 other contaminants, e.g., X-ray emission from the jet and/or corona. 
Accordingly, an exact measurement of the skewness will be difficult to obtain. 
However, we again note that if a black hole eclipse occurs in a binary system,
 an asymmetric light curve is expected for small accretion rates. 
Since the disk gas rotates at relativistic speeds around a black hole, 
 then the emitted radiation must be beamed by the Doppler effect. 

\section{Conclusion}

We have shown that it should be possible to observe the relativistic effects
 of a black hole using the expected asymmetry of its light curve at eclipse. 
Specifically, we propose the use of skewness and kurtosis analysis as  
 indicators of relativistic effects. 
In particular, we predict a skewness of $S \sim 0.08$, 
 which can be compared with observations.  

If asymmetry is not detected in the light curves at eclipse, 
 one possibility is that the emission from the inner region of the accretion disk
 is hidden by the disk's outer region, i.e., a self-occultation effect
 by a geometrically thick disk. 
For this to occur, the accretion rate must be very high relative to
 the critical accretion rate. 
Therefore, our analysis technique can be reversed to 
 use the skewness measurements to constrain the accretion rate. 

Even if we consider neutron stars, an asymmetry should appear in the light curves. 
In some eclipsing neutron star binaries, X-ray eclipses have been clearly seen
 in their light curves (Homan et al. 2003). 
In general, the apparent size of a neutron star is an order of magnitude
 smaller than that of a black hole. 
Thus, the time scales of the eclipse light curve features will be short
 ($\Delta t_{\rm NS} \sim $ 0.1 $\Delta t_{\rm BH}$). 
However, the time scale also depends on the binary parameters, in particular,
 the orbital period. 
For an X-ray binary with a relatively long orbital period,
 the eclipse light curves can be analyzed with present X-ray telescopes, 
 i.e., {\it RXTE}, {\it XMM-Newton}, or {\it ASTRO-E2}.

Moreover, if high-speed photometry on timescales of 0.01-10 seconds becomes available, 
 our analysis can be applied in the optical band to further increase understanding
 of black holes. 
Investigations of black holes using high-speed photometry
 will provide an interesting challenge in the near future.

We are grateful to Drs. H. Negoro, K. Matsumoto, and A. Yonehara
 for doing useful comments and discussions. 
This research was partially supported by the Ministry of Education, Science, Sports and Culture,
 Grant-in-Aid for JSPS Fellows, (16004706, KW;  15052631, 17010519, RT).
This work was also supported in part by the Grants-in Aid of the
Ministry of Education, Science, Sports, and Culture of Japan
(15540235, JF).


%
\begin{table}
  \begin{center}
  \caption{Eclipsing Black Hole X-ray Binary Candidates. }
    \begin{tabular}{llllll}
      \hline
      \hline
      Object name  & Black Hole Mass [$M_{\odot}$] &  $P$ [day]  & $\Delta t$ [s]  & $t_{\rm eclipse}$ [s]  & References \\
      \hline
      GRS 1915+105 & $\sim 14$                     & $\sim$ 33.5 &  12.0             & 5.9 $\times 10^3$ (K-M)   & Greiner et al. (2001) \\
      GRO J1655-40 & $\sim 7$                      & $\sim$ 2.62 &  3.04             & 5.5 $\times 10^3$ (F3)   & Orosz \& Bailyn (1997) \\
      V4641 Sgr    & $\sim 9.6$                    & $\sim$ 2.82 &  3.56             & 1.1 $\times 10^4$ (B9)   & Orosz et al. (2001) \\
      M33 X-7      & $\sim 2.5$	                   & $\sim$ 3.45 &  0.78             & 4.6 $\times 10^4$ (O-B)   & Pietsch et al. (2004) \\
      \hline
    \end{tabular}
  \end{center}
\label{tab:first}
\end{table}
%



\end{document}